\documentclass[%
reprint,
superscriptaddress,
 amsmath,amssymb,
 aps,
]{revtex4-1}

\usepackage{graphicx}% Include figure files
\usepackage{dcolumn}% Align table columns on decimal point
\usepackage{bm}% bold math
\usepackage[mathlines]{lineno}% Enable numbering of text and display math
\usepackage[colorlinks=true]{hyperref}
\usepackage{mathrsfs}
\usepackage{xcolor}
\usepackage{float}
\usepackage{ulem}
\usepackage[utf8]{inputenc}
\newcommand{\be}{\begin{equation}}
\newcommand{\ee}{\end{equation}}
\newcommand{\bea}{\setlength\arraycolsep{2pt} \begin{eqnarray}}
\newcommand{\eea}{\end{eqnarray}}

\begin{document}

%\preprint{APS/123-QED}

\title{The fastest relaxation rate of higher-dimensional Reissner--Nordstr\"{o}m black hole}% Force line breaks with \\
\author{Ming Zhang}
\email{mingzhang@jxnu.edu.cn}
\affiliation{College of Physics and Communication Electronics, Jiangxi Normal University, Nanchang 330022, China}
\affiliation{Department of Physics, Beijing Normal University, Beijing 100875, China}
\author{Jie Jiang}
\email{iejiang@mail.bnu.edu.cn}
\affiliation{Department of Physics, Beijing Normal University, Beijing 100875, China}
\author{Zhen Zhong}
\email{Corresponding author: zhenzhong@mail.bnu.edu.cn}
\affiliation{Department of Physics, Beijing Normal University, Beijing 100875, China}

\date{\today}% It is always \today, today,

\begin{abstract}
In the eikonal regime, we analytically calculate quasinormal resonance frequencies for massless scalar perturbations of the higher-dimensional Reissner--Nordstr\"{o}m (RN) black holes. Remarkably, we find that the higher-dimensional RN black holes coupled with the massless scalar fields have the fastest relaxation rates in the Schwarzschild limit, this is qualitatively different from the four-dimensional case where the black hole with non-vanishing charge has the fastest relaxation rate.
%\begin{description}
%04.70.Dy, 04.70.-s
%\end{description}
\end{abstract}

%\pacs{04.70.Dy, 04.70.-s}% PACS, the Physics and Astronomy

\maketitle

%\tableofcontents

\section{Introduction}%\label{Introduction}
The quasinormal mode, which was detected recently in form of gravitational wave signals \cite{Abbott:2016blz,TheLIGOScientific:2016src,Abbott:2016nmj} and has been investigated for about five decades since the seminal work \cite{Regge:1957td}, is a fascinating object of research in gravitational physics. The quasinormal modes are in fact the black hole's reactions to perturbations. Recent interests related to it can be seen in \cite{Cardoso:2017soq,Destounis:2018qnb,Gwak:2018rba,Ge:2018vjq,Luna:2018jfk,Dias:2018ufh,Mo:2018nnu,Cardoso:2018nvb,Dias:2018etb,Hod:2018lmi,Dias:2018ynt,Hod:2018dpx} for strong cosmic censorship  (especially for higher-dimensional cases, e.g., \cite{Liu:2019lon,Rahman:2018oso}) and in \cite{Hod:2017rmh,Hod:2015hza,Hod:2016bas,Hod:2016vkt,Hod:2018fet,Hod:2017www,Hod:2017kpt,Huang:2017whw,Huang:2015cha,Li:2019tns} for black hole's no-hair theorem (especially for higher-dimensional cases, e.g., \cite{Zhang:2018jgj}).

The uniqueness theorem \cite{Israel:1967wq,Carter:1971zc,Hawking:1971vc,Robinson:1975bv,Hollands:2012vls,Ganchev:2017uuo,Barcelo:2019aif} tells us that curved spacetime described by Kerr-Newman(KN) metric in the Einstein-Maxwell gravitational theory can hold at most three global conserved quantities, which are named as mass, electric charge and angular momentum. Correspondingly, no-hair conjecture \cite{Ruffini:1971bza,Nunez:1996xv,Torii:1998ir,Hod:2012px,Hod:2017rmh,Hod:2017kpt,Bhattacharjee:2017huz,Costa:2018zvw} for black holes states that static matter field (such as the massless scalar field) perturbations of the black hole in KN family can not be permanently hold. It means that once physically proper boundary conditions, which yield purely ingoing waves at the event horizon of the black hole and purely outgoing waves at spatial infinity, are set, the quasinormal resonant modes describing the canonical family of KN black hole spacetimes against the matter field perturbations are characterized with decaying frequencies \cite{Nollert:1993zz,Hod:1998vk,Kokkotas:1999bd,Nollert:1999ji,Berti:2009kk,Konoplya:2011qq}. These decaying modes lead to the absorbing and scattering of the perturbations by the black hole, resulting in a stationary or static black hole in KN family.

The boundary conditions at the event horizon of the black hole and the spatial infinity can determine a set of discrete quasinormal resonance spectrum functions, which can be denoted as $\{\omega_{n}(M,Q\cdots)\}_{n=0}^{n=\infty}$, with $M,Q$ respectively the mass and electric charge of the black hole \cite{Schutz:1985zz,Iyer:1986np,Konoplya:2003ii,Matyjasek:2017psv,Konoplya:2019hlu,Santos:2019yzk}. The imaginary part of the fundamental resonant frequency $\omega_{0I}\equiv\omega_{I}(n=0)$ embodies information about the characteristic relaxation time of the composed perturvation field/black hole system by the relation
\begin{equation}
\tau=\omega^{-1}_{0I}.
\end{equation}

For the charged massive scalar perturbation of $d$-dimensional Reissner-Nordstr\"{o}m (RN) black hole, we have obtained that $\tau$ becomes infinite under a condition $\left[2(d-3)\mu M\right]/\left[(d-2)qQ\right]\to 1^{-}$ ($\mu$ and q are individually the mass and electric charge of the scalar field) in \cite{Zhang:2018jgj} based on \cite{Hod:2016jqt}, which means that the relaxation rate can be zero in that extreme condition. For the massless scalar perturbation of four-dimensional Kerr or KN  black hole, it has been shown that the relaxation rate will decrease to be vanishing once $MT\to 0$ ($T$ is the temperature of the black hole) \cite{Hod:2008zz,Hod:2008se,Hod:2012bw,Berti:2005eb,Pani:2013wsa,Hod:2014uqa,Dias:2015wqa}.

In \cite{Hod:2018ifz}, relaxation rate of the four-dimensional RN black hole coupled with massless scalar field was investigated. Recognizing that the maximized relaxation rate of the perturbed black hole in KN family can be reached by considering a non-rotating one (as the relaxation rate decreases with the increasing angular momentum per unit mass $a$ \cite{Hod:2015xlh}), \cite{Hod:2018ifz} analytically gives the relaxation time for the composed massless scalar field/four-dimensional RN black hole system in the eikonal regime $l\gg 1$ ($l$ is the spherical harmonic index). Moreover, it was found that the relaxation rate is maximized when $Q/M\sim 0.7$. This result is consistent with numerically calculated ones \cite{Pani:2013wsa,Pani:2013ija}.

 Instead of concerning on the massless scalar field perturbation for the four-dimensional RN black hole, we here will extend the work in \cite{Hod:2018ifz} to higher-dimensional case. Intuitively, we may think that the dimensions of the spacetime do not qualitatively change the result in \cite{Hod:2018ifz}. However, in the remaining part of the paper we will show that the dimension of the spacetime has an extraordinary effect on the relaxation rate. In Sec. \ref{chapter2}, we will introduce the massless scalar perturbation equation for the higher-dimensional RN black hole. In Sec. \ref{chapter3}, we will analytically derive the quasinormal resonance frequency of the composed massless scalar field/higher-dimensional RN black hole system. In Sec. \ref{chapter4}, we will give some closing remarks.

\section{Massless scalar field perturbation equation of higher-dimensional RN black hole}\label{chapter2}
Using the natural units with $G=c=\hbar=1$, the $d$-dimensional RN black hole can be described by the line element
\begin{equation}
ds^2=-f(r)dt^2+\frac{dr^2}{f(r)}+r^2 d\Omega^2_{d-2},
\end{equation}
where
\begin{equation}
f(r)=1-\frac{16 \pi  M r^{3-d}}{(d-2) \Omega_{d-2} }+\frac{32 \pi ^2 Q^2 r^{-2 (d-3)}}{\left(d^2-5 d+6\right) \Omega_{d-2} ^2}.
\end{equation}
Here $\Omega_{d-2}= 2 \pi ^{\frac{\text{d}-1}{2}}/\Gamma \left[(d-1)/2\right]$ is the volume of a $(d-2)$-sphere. The electromagnetic field $F$ and gauge potential $A$ are
\begin{equation}
  F_{ab}=(dA)_{ab},~A_a=\frac{4 \pi  Q r^{3-d}}{(3-d)\Omega_{d-2}}(dt)_{a}.
\end{equation}
The Klein-Gordon wave equation describing the propagation of a massless scalar field $\Psi (t,r,\Theta)$ in higher-dimensional spacetime reads
\begin{equation}\label{kg}
\triangledown^{a}\triangledown_{a}\Psi(t,r,\Theta)=0.
\end{equation}
Decomposing the scalar field function by spherically harmonic function through
\begin{equation}
  \Psi (t,r,\theta)=\int\sum_{lm}e^{-i\omega t}\frac{R_{lm}(r, \omega)}{r^{\frac{d-2}{2}}}Y_{lm}(\theta),
\end{equation}
where $m$ is the azimuthal harmonic index, and substituting it into the Klein-Gordon wave equation (\ref{kg}), we can obtain differential equations of $R_{lm}(r, \omega)$ describing the radial dynamics  and $Y_{lm}(\theta)$ describing the angular dynamics for higher-dimensional black hole against the massless scalar field perturbation. The eigenvalue that the radial and angular components share is $K_l =l(d+l-3)$. Explicitly, the master radial equation can be written as
\begin{equation}\label{radial}
 f(r)^{2} R''(r)+f'(r)f(r) R'(r)+U R(r)=0,
\end{equation}
where
\begin{equation}
  U=\omega^2-\frac{(d-4) (d-2) f(r)^2}{4 r^2} -\frac{f(r) \left[(d-2) r f'(r)+2 K_l\right]}{2 r^2}
\end{equation}
and ${}^{\prime}$ denotes derivative with respect to the coordinate $r$.

Taking advantage of a tortoise coordinate $r_{*}$, which is defined by a differential relation
\begin{equation}
\text{d}r_{*}=\frac{\text{d}r}{f(r)},
\end{equation}
the radial equation (\ref{radial}) can be reduced to an ordinary differential equation
\begin{equation}\label{sch}
\frac{\text{d}^2 R}{\text{d}r_{*}^2}+(\omega^2 -V)R=0,
\end{equation}
which is Schr\"{o}dinger-like. The radial effective potential $V$ can be expressed as
\begin{equation}\label{pote}
 V=f(r) H(r),
\end{equation}
where
\begin{equation}
\begin{aligned}
H(r)=&\frac{K_l}{r^2}+\frac{d^2-6 d+8}{4 r^2}\\&+\frac{4 \pi  (d-2) M}{\Omega_{d-2}  r^{d-1}}-\frac{8 \pi ^2 (3 d-8) Q^2}{(d-3) \Omega_{d-2} ^2 r^{2 d-4}}.
\end{aligned}
\end{equation}
The radial wave at the outer horizon $r_{+}$ of the higher-dimensional RN black hole must be purely ingoing, and the radial wave at the spatial infinity must be purely outgoing. That is to say, the condition
\begin{equation}\label{boundaryc}
R \backsim\left\{
\begin{aligned}
& e^{-i\omega r_{*}},~~~~~r\to r_+ ~(r_{*}\to -\infty); \\
& e^{i\omega r_{*}},~~~~~~~r\to \infty ~(r_{*}\to \infty)
\end{aligned}
\right.
\end{equation}
should be affiliated to (\ref{sch}). As a result, we can single out the quasinormal resonant modes $\left\{\omega _n(M,Q,\mu ,q,l)\right\}_{n=0}^{n=\infty }$ characterizing the relaxation dynamics of the massless scalar fields perturbating the $d$-dimensional RN black hole.

\section{The quasinormal resonance frequency of the composed  massless scalar field/higher-dimensional RN black hole system}\label{chapter3}
The event horizon radius $r_{+}$ of the $d$-dimensional RN black hole can be expressed as
\begin{equation}
\begin{aligned}
r_+ =&\left(\frac{4 \pi}{\Omega_{d-2}} \right)^{\frac{1}{d-3}}\\&\times\left(\frac{2 M}{d-2}+\sqrt{\frac{4 M^2}{(d-2)^2}-\frac{2 Q^2}{d^2-5 d+6}}\right)^{\frac{1}{d-3}},
\end{aligned}
\end{equation}
from which we can obtain 
\begin{equation}\label{evcon}
M^2 \geqslant \frac{(d-2) Q^2}{2 (d-3)}.
\end{equation}
Then at the position $r$ outside the event horizon, we have
\begin{equation}
\begin{aligned}
r>r_+ &\geqslant \left[\frac{8 \pi  M}{(d-2) \Omega_{d-2} }\right]^{\frac{1}{d-3}}\\&\geqslant\left[\frac{8\pi Q}{\Omega_{d-2}\sqrt{2(d-2)(d-3)}}\right]^{\frac{1}{d-3}}.
\end{aligned}
\end{equation}
As a result, we can know
\begin{equation}
\frac{4 \pi  (d-2) M}{\Omega_{d-2}  r^{d-3}}<\frac{(d-2)^{2}}{2},
\end{equation}
\begin{equation}
\frac{8 \pi ^2 (3 d-8) Q^2}{(d-3) \Omega_{d-2} ^2 r^{2 d-6}}<\frac{(d-2)(3d-8)}{4}.
\end{equation}
In the eikonal regime,
\begin{equation}\label{eik}
l\gg 1,~K_{l}\sim l^{2}\gg 1,
\end{equation}
we further have
\begin{equation}\begin{aligned}
H(r)&=\frac{K_{l}}{r^{2}} \left[1+\frac{d^2-6 d+8}{4 K_{l}}\right.\\&~~~~+\left.\frac{4 \pi  (d-2) M}{\Omega_{d-2}  r^{d-3}K_{l}}-\frac{8 \pi ^2 (3 d-8) Q^2}{(d-3) \Omega_{d-2} ^2 r^{2 d-6}K_{l}}\right]\\&\sim \frac{K_{l}}{r^{2}}.
\end{aligned}\end{equation}

\begin{figure*}[!ht] %  figure placement: here, top, bottom, or page
   \centering
   \includegraphics[width=2.6in]{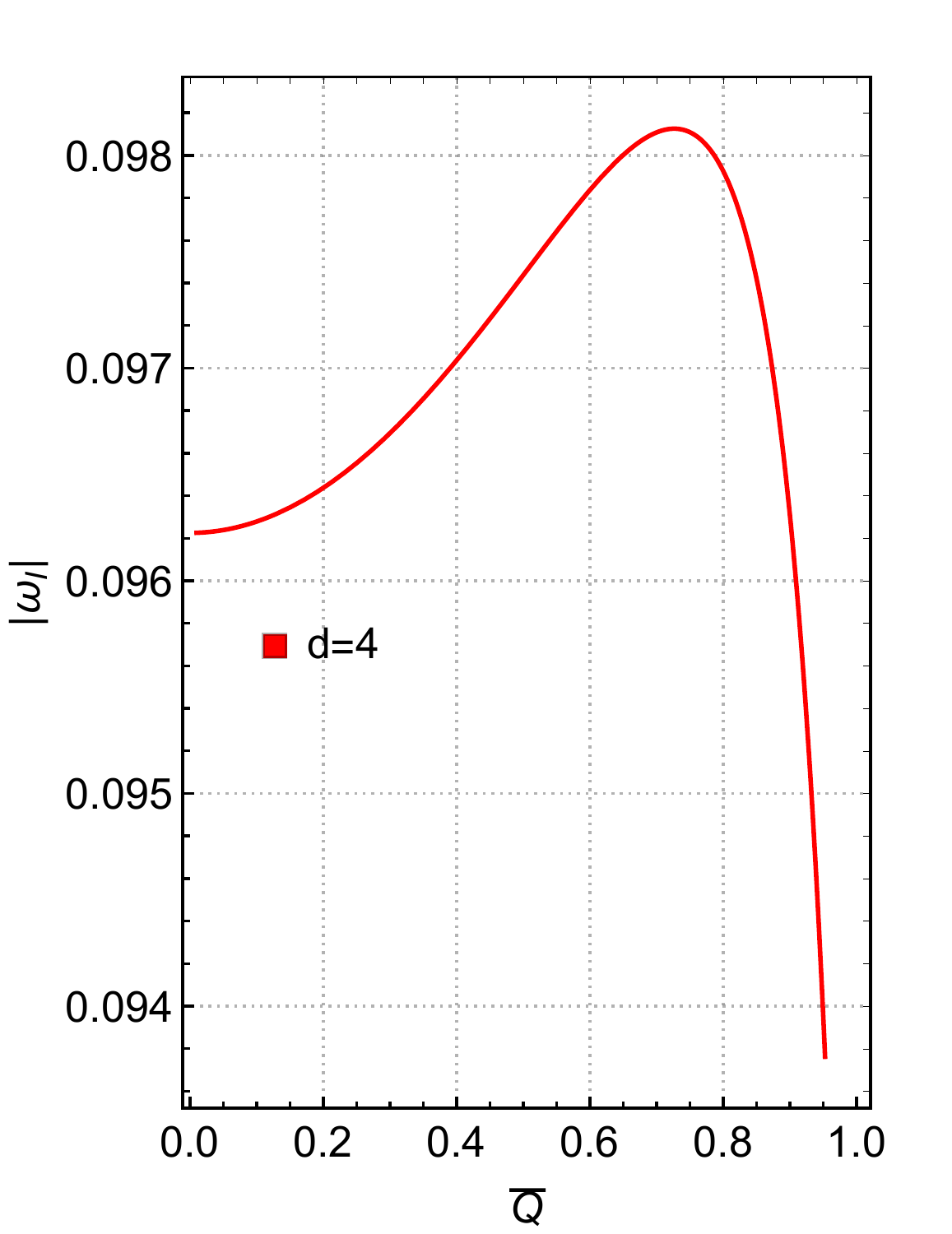}
    ~~~~~\includegraphics[width=2.48in]{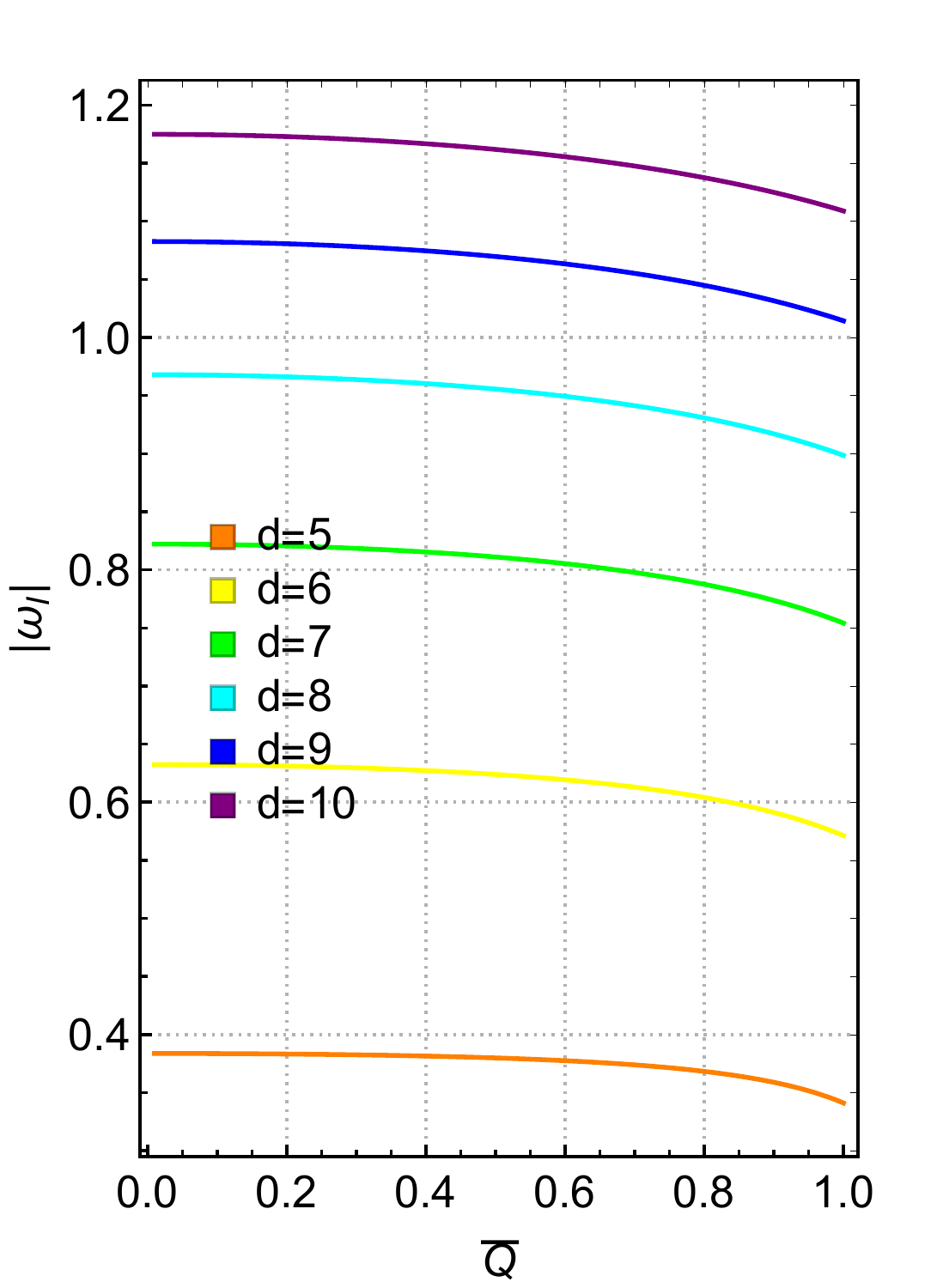}
   \caption{The variation of the imaginary part $\omega_I$ of the quasinormal resonance frequency with respect to $\bar{Q}$ for $M=1,~n=0$.}
   \label{lambdadiagram}
\end{figure*}

\begin{table*}[!htb]
\centering
\caption{The fastest relaxation rate of the composed scalar field/higher-dimensional RN black hole system for $M=1,~n=0$.}
\begin{tabular}{ccccccc} % ???????
  \hline\hline
$~~~~~~~d~~~~~~~$ &~~~~~~~5~~~~~~~&~~~~~~~6~~~~~~~&~~~~~~~7~~~~~~~&~~~~~~~8~~~~~~~&~~~~~~~9~~~~~~~&~~~~~~~10\\
\hline
$~~~~~~~\bar{\omega_{I}}(\bar{Q}\to 0)~~~~~~~$ &~~~~~~~0.3837~~~~~~~&~~~~~~~0.6324~~~~~~~&~~~~~~~0.8222~~~~~~~&~~~~~~~0.9679~~~~~~~&~~~~~~~1.0826~~~~~~~&~~~~~~~1.1748\\
\hline
\hline
\end{tabular}
  \label{tbl:table1}
\end{table*}

Then the radial effective potential  can be approximately written as
\begin{equation}\label{appveff}
V=\frac{K_{l}f(r)}{r^{2}}.
\end{equation}

The radial effective potential $V$ in the Schr\"{o}dinger-like ordinary differential equation (\ref{pote})  acts as an effective potential barrier. By the derivative with respect to the coordinate $r$ for the approximate radial effective potential (\ref{appveff}), we can obtain the effective potential's maximum value at the point
\begin{equation}
r_0=\left[\frac{4 \pi  (d-1) M}{(d-2) \Omega_{d-2}}\left[1+\sqrt{1-\frac{2 (d-2)^2 Q^2}{(d-3) (d-1)^2 M^2}}\right]\right]^{\frac{1}{d-3}}.
\end{equation}

The quasinormal resonance frequency of the composed massless scalar field/higher-dimensional RN black hole system can be got by the WKB resonance condition \cite{Iyer:1986np}
\begin{equation}\label{wkb}
  \frac{\mathcal{Q}_0}{\sqrt{2\mathcal{Q}_0^{(2)}}}=-i\left(n+\frac{1}{2}\right)+\mathcal{O}\left(\frac{1}{l}\right),
\end{equation}
where $n$ is the overtone number,
\begin{equation}
\mathcal{Q}_{0}\equiv\omega^2 -V(r=r_{0})
\end{equation}
and
\begin{equation}
\mathcal{Q}_0^{(2)}\equiv \left.\frac{\text{d}^2 \mathcal{Q}[r(r_{*})]}{\text{d}r_{*}^{2}}\right |_{r=r_{0}}.
\end{equation}

\begin{figure*}[!ht] %  figure placement: here, top, bottom, or page
   \centering
   ~\includegraphics[width=4.5in]{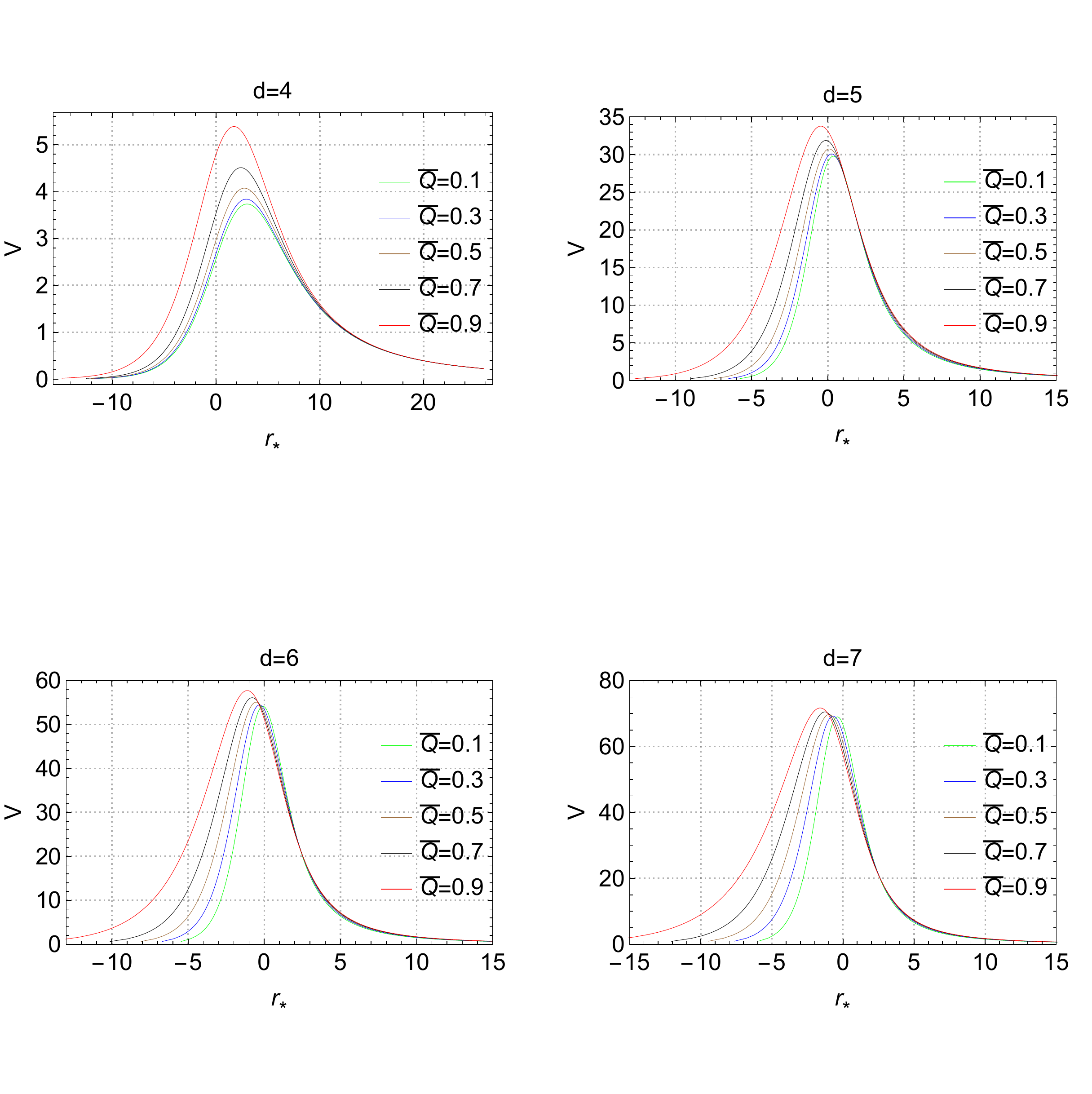}
   \caption{The variation of the effective potential with respect to the tortoise coordinate $r_*$ for $M=1, l=100$.}
   \label{eff}
\end{figure*}

\begin{figure*}[!ht] %  figure placement: here, top, bottom, or page
   \centering
    \includegraphics[width=5.5in]{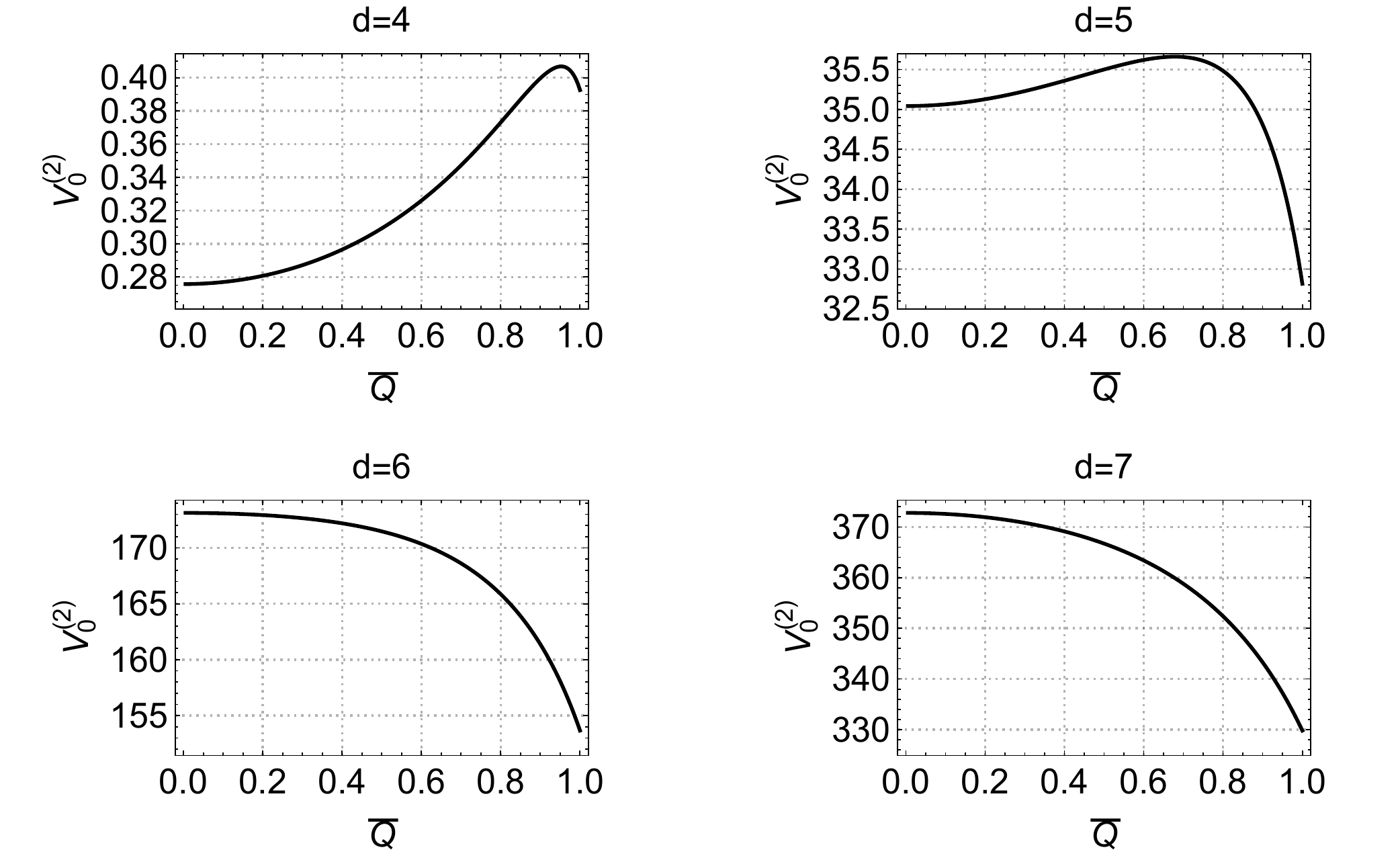}
   \caption{The variation of $V_0^{(2)}$ with respect to $\bar{Q}$ for $M=1, l=100$.}
   \label{effd}
\end{figure*}

In order to analytically solve (\ref{wkb}), we here consider the eikonal regime with large spherical harmonic index $l$, which leads to an approximation expression (\ref{appveff}) for the radial effective potential (\ref{pote}). Additionally, we require that the real part of the quasinormal resonance frequency $\omega_{R}$ be much bigger than the imaginary part of the quasinormal resonance frequency $\omega_{I}$, {\it{i.e.}},
\begin{equation}\label{rbiggeri}
\omega_R\gg \omega_I.
\end{equation}

Furthermore, (\ref{wkb}) can be expressed as
\begin{equation}\label{reso}
 \frac{\omega^2-\frac{K_{l}f(r_{0})}{r_{0}^{2}}}{f(r_{0})\sqrt{2V_{0}^{\prime\prime}}}=-i(n+\frac{1}{2}).
\end{equation}
The real part of the quasinormal resonance mode for the composed scalar field/black hole system can be obtained as
\begin{equation}
\omega_R=\frac{\sqrt{K_{l}f(r_{0})}}{r_{0}}\sim \frac{l\sqrt{f(r_{0})}}{r_{0}},
\end{equation}
by solving
\begin{equation}
\omega_{R}^2-\frac{K_{l}f(r_{0})}{r_{0}^{2}}=0.
\end{equation}
Solving
\begin{equation}
  2 \omega _I \omega_R=\left(n+\frac{1}{2}\right) f(r_0)\sqrt{2V_{0}^{\prime\prime}}
\end{equation}
yields the imaginary part of the quasinormal resonance frequency
\begin{equation}
\omega_I=\frac{\sqrt{2}(2n+1)r_{0}\sqrt{f(r_{0})V_{0}^{\prime\prime}}}{4\sqrt{K_{l}}}.
\end{equation}

A dimensionless quantity
\begin{equation}
\bar{Q}\equiv\frac{\sqrt{d-2}}{\sqrt{2(d-3)}}\frac{Q}{M}.
\end{equation}
can be defined and we can know
\begin{equation}
0\leqslant\bar{Q}\leqslant 1
\end{equation}
by considering (\ref{evcon}), where $\bar{Q}=0$ corresponds to the RN black hole's Schwarzschild limit and $\bar{Q}=1$ corresponds to extreme RN black hole. Then we can simplify expressions of the quasinormal resonance frequencies. We here show some explicit imaginary parts of the modes with $M=1$.  For $d=4$, 
\begin{equation}
\omega_{I}(d=4)=\frac{\left(4 \bar{Q}^2+x_1-3\right) \sqrt{\frac{-\left(33-7 x_1\right) \bar{Q}^2+8 \bar{Q}^4-9 \left(x_1-3\right)}{1-\bar{Q}^2}}}{32 \bar{Q}^4},
\end{equation}
with
\begin{equation}
x_{1}=\sqrt{9-8 \bar{Q}^2}.
\end{equation}
This is consistent with the result in \cite{Hod:2018ifz}. For $d=5,~6$, we have
\begin{equation}
\begin{aligned}
\omega_{I}(d=5)=&\frac{1}{3} \sqrt{\frac{\pi }{6}} \left(3 \bar{Q}^2+x_2-2\right)\\&\times \sqrt{\frac{\left(x_2-3\right) \bar{Q}^2-2 x_2+4}{\bar{Q}^6-\bar{Q}^8}},
\end{aligned}
\end{equation}
and
\begin{equation}
\omega_{I}(d=6)=\frac{3^{2/3} \sqrt[3]{\pi } \sqrt{\left(25x_3-8 \left(x_3-5\right) \bar{Q}^2\right) \left(8 \bar{Q}^2+x_3\right)}}{16\ 2^{5/6} \sqrt[6]{5-x_3} \bar{Q}^{8/3}},
\end{equation}
where
\begin{equation}
x_{2}=\sqrt{4-3 \bar{Q}^2},
\end{equation}
\begin{equation}
x_{3}=\sqrt{25-16 \bar{Q}^2}-5.
\end{equation}

\section{Closing remarks}\label{chapter4}
\cite{Hod:2018ifz} gives an answer to the question that which black hole in the canonical family of Kerr-Newman spacetimes coupled with massless scalar field  owns the fastest relaxation rate. It was analytically found that the Kerr-Newman black hole with zero angular momentum and $\bar{Q}\sim 0.7$ has the least relaxation time in the eikonal regime where the spherical harmonic index $l\gg 1$.  In this paper, we further think about the relaxation rate problem in higher-dimensional spacetime. We wonder that whether or not the property of the relaxation rate for higher-dimensional RN black hole coupled with massless scalar field behaves like the four-dimensional counterpart.

First, we have shown the perturbation equation for the higher-dimensional RN black hole coupled with massless scalar field, together with physically reasonable boundary conditions both at the event horizon and spatial infinity. Then, in the eikonal limit $l\gg 1$, we have given an approximate expression for the effective potential of the composed massless scalar field/higher-dimensional black hole system. After that, to single out the quasinormal resonance frequency of the system, we have used the WKB method. Supposing that the imaginary part is much smaller than the real part for the quasinormal resonance frequency, we  have analytically obtained the modes, which can be reduced to the result in \cite{Hod:2018ifz} describing the four-dimensional case. The obtained quasinormal modes as  resonance solutions of the Schr\"{o}dinger-like equation (\ref{sch}) can be expressed as
\begin{equation}
\omega=\frac{l\sqrt{f(r_{0})}}{r_{0}}-i\frac{\sqrt{2}(2n+1)r_{0}\sqrt{f(r_{0})V_{0}^{\prime\prime}}}{4\sqrt{l(d+l-3)}}.
\end{equation}\label{ome}

For the composed massless scalar field/four-dimensional RN black hole system, one can see from the left diagram in Fig. \ref{lambdadiagram} that the relaxation rate first increases with the increasing dimensionless quantity $\bar{Q}$ and then climbs up to a maximum value for $\bar{Q}\sim 0.7$ before decreasing steeply.

For the composed massless scalar field/higher-dimensional RN black hole system, one can see from the right diagram in Fig. \ref{lambdadiagram} that the relaxation rates just monotonously decrease from their maximum values at $\bar{Q}=0$ (which corresponds to the higher-dimensional RN black hole's Schwarzschild limit) to their minimum values at $\bar{Q}=1$ (which corresponds to the extreme RN black hole). This is qualitatively different from the four-dimensional case. We have listed the value of the imaginary part $\omega_{I}$ of the dominant fundamental quasinormal resonance frequency ($n=0$) in the RN black hole's Schwarzschild limit $\bar{Q}\to 0$ in Table \ref{tbl:table1}. The result shows that higher-dimensional black hole perturbation \cite{Konoplya:2003ii,Cardoso:2002pa,Konoplya:2003dd,Cardoso:2003vt,Cardoso:2003qd,Konoplya:2008rq,Chakrabarti:2008xz,Zhidenko:2006rs,Cornell:2005ux,Moderski:2005hf,Cardoso:2004cj} have characteristics different from the common four-dimensional one.

The difference between four-dimensional relaxation rate and higher-dimensional ones stems from characteristics of the effective potentials for the composed massless scalar field/black hole system as
\begin{equation}
\tau^2\sim \frac{V_0}{V_0^{(2)}}.
\end{equation}
We show the effective potential in Fig. \ref{eff} and the effective potential's second-order derivative at the extreme point $r_0$ in Fig. \ref{effd}. From Fig. \ref{eff}, we can see that the shapes of the effective potentials seem alike for different spacetime dimensions and the extreme values of the effective potentials increase with increasing electric charge for all spacetime dimensions. This can not explain why there are different relaxation rate pattens between four-dimensional system and higher-dimensional system. What makes sense can be found in Fig. \ref{effd}, where we can see that, for the four-dimensional composed system, the second-order derivative of the effective potential at the extreme point increases with the increasing electric charge, but the value begins to decrease when the black hole becomes nearly extreme; for the five-dimensional system, the curve behaves similarly with the four-dimensional one, but the increasing part seems to be gentler and the turning point, after which the value decreases, appears earlier; for the spacetime dimensions larger than five, the value of $V_0^{(2)}$ decreases monotonically for increasing $\bar{Q}$. Considering the behaviors of the effective potentials in the diagrams, we can understand that it is the second-order derivative of the effective potential at the peak that makes the relaxation rates between the four-dimensional and higher-dimensional composed  massless scalar field/black hole systems different.

Our result is in consistent with the investigation in  \cite{Andersson:1996xw}, where it was found that the oscillation mode of the composed massless scalar field/higher-dimensional RN black hole system mainly  depends on the the maximal height of the effective potential at the extreme point $r_0$, whereas the slowly damping mode of the system is mainly affected by the second-order derivative of the effective potential at the extreme point $r_0$.

In fact, the quasinormal mode of the composed massless scalar field/spherically symmetric black hole system in the eikonal limit is related to the circular geodesic motion of the null particle surrounding the black hole. The extreme point $r_0$, where the effective potential gets its maximum value, is just the radius of the null circular geodesic $r_c$  \cite{Cardoso:2008bp}. Correspondingly, the quasinormal frequency for higher-dimensional RN black hole against massless scalar field can be expressed as \cite{Cardoso:2008bp,Decanini:2010fz}
\begin{equation}
\omega=\Omega_c l-i \left(n+\frac{1}{2}\right)|\lambda|,
\end{equation}
where $\Omega_c$ is the coordinate angular velocity of the circular null geodesics, $\lambda$ is the Lyapunov exponent. These two quantities are respectively related with the orbital time scale $T_\Omega$ and instability time scale $T_\lambda$ by
\begin{equation}
T_\Omega \equiv \frac{2\pi}{\Omega_c},~T_\lambda\equiv\frac{1}{|\lambda|}.
\end{equation}
Our finding shows that, considering the variation of the black hole charge, the instability time scales of the null particles around the higher-dimensional black holes are qualitatively different from that of the four-dimensional case.

For the four-dimensional black hole in the canonical Kerr-Newman spacetimes, it was found that the one with vanishing rotation has maximal relaxation rate \cite{Hod:2008se,Pani:2013wsa,Dias:2015wqa} against the massless scalar field. One may wonder that whether this result is applicable to the canonical family of higher-dimensional black hole in the Einstein gravity. (Specifically, they are higher-dimensional (charged) Schwarzschild black hole and Myers-Perry black holes \cite{Myers:1986un}. Though accelerating black holes \cite{Anabalon:2018ydc,Anabalon:2018qfv,Zhang:2019vpf} are also solutions in Einstein theory, there are no higher-dimensional counterparts found yet.) The answer, according to the known work on this, may be positive. Quasinormal mode of the higher-dimensional rotating Kerr black hole with a single axis against massless scalar field was studied, especially for the five-dimensional case \cite{Ida:2002zk} (where there is an upper bound for the rotation parameter, just like the four-dimensional one) and the six-dimensional case \cite{Cardoso:2004cj} (where there are no upper bounds for the rotation parameters and there are no extreme black holes). It was found that the higher dimensional rotating Kerr black holes are stable against massless scalar field perturbations. In \cite{Ida:2002zk}, it was shown that the imaginary part of the frequency is negative for the five-dimensional rotating Kerr black hole against massless scalar field. According to the numerical results shown in \cite{Cardoso:2004cj}, we can know that, for the fundamental mode with $j=1$ ($j$ is related to the separation eigenvalue of the radial and angular equations), the imaginary part of the quasinormal frequency first decreases and then slightly increases before finally asymptotically approaching to zero, which seems to indicate that the fastest relaxation rate can be obtained for a higher-dimensional rotating black hole with vanishing rotation. Given all that, we suspect that the higher-dimensional black hole in the Einstein gravity against the massless scalar field may get the fastest relaxation rate if its charge or rotation vanishes. However, more investigations are needed to be done with the scalar field perturbations (also gravitational perturbation \cite{Kunduri:2006qa,Kodama:2009bf,Murata:2008yx}) of higher-dimensional Kerr black holes (with more than one rotating axes).

\section*{Acknowledgements}
The authors thank Hongbao Zhang for his useful discussions. This work is supported by the National Natural Science Foundation of China (NSFC) with Grants No. 11775022, 11375026, 11475179 and 11675015. 

\bibliography{Notes}

\end{document}